\documentclass[11pt,onecolumn]{article}
\usepackage{amsfonts}

\input{tcilatex}

\begin{document}

\title{Dynamical Structure of Irregular Constrained Systems}
\author{Olivera Mi\v {s}kovi\'{c}$^{*,\dagger }$ and Jorge Zanelli$^{*}\medskip $ \\
$^{*}$\emph{Centro de Estudios Cient\'{i}ficos (CECS), Casilla 1469,
Valdivia, Chile.}\\
$^{\dagger }$\emph{Departamento de F\'{i}sica, Universidad de Santiago de
Chile, }\\
\emph{Casilla 307, Santiago 2, Chile.}}
\date{}
\maketitle

\begin{abstract}
Hamiltonian systems with functionally dependent constraints (irregular
systems), for which the standard Dirac procedure is not directly applicable,
are discussed. They are classified according to their behavior in the
vicinity of the constraint surface into two fundamental types. If the
irregular constraints are multilinear (type I), then it is possible to
regularize the system so that the Hamiltonian and Lagrangian descriptions
are equivalent. When the constraints are power of a linear function (type
II), regularization is not always possible and the Hamiltonian and
Lagrangian descriptions may be dynamically inequivalent. It is shown that
the inequivalence between the two formalisms can occur if the kinetic energy
is an indefinite quadratic form in the velocities. It is also shown that a
system of type I can evolve in time from a regular configuration into an
irregular one, without any catastrophic changes. Irregularities have
important consequences in the linearized approximation to nonlinear
theories, as well as for the quantization of such systems. The relevance of
these problems to Chern-Simons theories in higher dimensions is discussed.
\end{abstract}

\section{Introduction}

Dirac's Hamiltonian analysis provides a systematic method for finding the
gauge symmetries and the physical degrees of freedom of constrained systems
like gauge theories and gravity \cite{Dirac}. Constraints arise through
conditions of the form 
\begin{equation}
\phi ^{r}(z)\approx 0\qquad (r=1,...,R),  \label{constraints}
\end{equation}
where $z$ are local coordinates in phase space $\Gamma $. In the most common
cases of physical interest the $\phi $'s are functionally independent; these
are the regular constrained systems. There are some exceptional cases in
which functional independence is violated. In these \emph{irregular systems}
it is not always clear how to identify symmetries and true degrees of
freedom. Moreover, the Hamiltonian and Lagrangian descriptions may not be
equivalent in irregular systems.

Irregular systems are not necessarily intractable nor exotic. A common
example is a relativistic massless particle ($p^{\mu }p_{\mu }=0$), which is
irregular at the origin of momentum space ($p^{\mu }=0$). There are other
physical circumstances in which regularity is violated, and not only for
isolated states but on large portions of phase space where the system
evolves. This is the case in Chern-Simons (CS) theories for dimensions $%
D\geq 5$ where, for some initial configurations, regularity can fail at all
times and one is forced to live with this problem.

A CS Lagrangian describes a gauge theory for a certain Lie group $G$ in a
spacetime of odd dimension. The construction is naturally invariant under
diffeomorphisms and provides a non-standard but otherwise acceptable
description of gravity as a gauge theory \cite{Chamseddine}-\cite
{Troncoso-Zanelli}. Furthermore, CS theories are highly nonlinear, possess
propagating degrees of freedom \cite{Banados-Garay-Henneaux}, and have a
very rich phase space structure with many different sectors, some of which
describe irregular systems \cite{Miskovic-Zanelli}.

In five-dimensional CS supergravity, it was observed that the linearized
action around a certain anti-de Sitter background seems to have \emph{one
more degree of freedom }than the fully nonlinear system \cite
{Chandia-Troncoso-Zanelli}. This paradoxical behavior can be seen to arise
from a violation of the regularity conditions among the symmetry generators
of the theory in the region of phase space defined by the selected
background.

Here we address different scenarios in which regularity conditions can be
violated, how such systems can be handled in some cases, and why
linearization may fail to approximate a nonlinear system faithfully. It is
found that there are two main types of irregular constraints, multilinear
(type I) and nonlinear (type II). The constraints of the first type can
always be regularized, while type II constraints (of the form $f^{k}$, where 
$f$ has a simple zero and $k>1$) can be regularized only if $f$ is a second
class function.

Constraints satisfying regularity conditions are sometimes referred to as 
\emph{effective} constraints \cite{Batlle-Gomis-Pons-RomanRoy}. The issue of
regularity (\emph{effectiveness}) and its relevance for the equivalence
between the Lagrangian and Hamiltonian formalisms has also been discussed in 
\cite{Garcia-Pons,Pons-Salisbury-Shepley}.

\section{Regularity conditions}

Consider a dynamical system in a phase space $\Gamma $ with local
coordinates $z^{i}\equiv (q,p)$ $\left( i=1,\ldots ,2n\right) $. Conditions (%
\ref{constraints}) define the constraint surface 
\begin{equation}
\Sigma =\left\{ \bar{z}\in \Gamma \mid \phi ^{r}(\bar{z})=0\;\left(
r=1,\ldots ,R\right) \left( R\leq 2n\right) \right\} .
\end{equation}
Dirac's procedure guarantees that the system remains on the constraint
surface during its evolution (for reviews, see Refs. \cite{Dirac},\cite
{Hanson-Regge-Teitelboim}-\cite{Chitaia-Gogilidze-Surovtsev}). Choosing
different coordinates on $\Gamma $ may lead to different forms for the
constraint functions whose functional independence is not obvious. The \emph{%
regularity conditions} were designed by Dirac to test this \cite{DiracCJM}%
.\medskip

\textbf{Regularity conditions (RCs)}: \emph{\ The constraints $\phi
^{r}\approx 0$ are regular if and only if their small variations $\delta
\phi ^{r}\ $evaluated on $\Sigma $ define $R$ linearly independent functions
of $\delta z^{i}$.}\medskip

To first order in $\delta z^{i}$, the variations of the constraints have the
form 
\begin{equation}
\delta \phi ^{r}=J_{i}^{r}\delta z^{i}\qquad (r=1,...,R),
\label{linearized constraints}
\end{equation}
where $J_{i}^{r}\equiv \left. \frac{\partial \phi ^{r}}{\partial z^{i}}%
\right| _{\Sigma }$ is the Jacobian evaluated on the constraint surface. An
equivalent definition of the RCs is \cite{Henneaux-Teitelboim}: \emph{The
set of constraints} $\phi ^{r}\approx 0$ \emph{is regular if and only if the
Jacobian} $J_{i}^{r}=\left. \frac{\partial \phi ^{r}}{\partial z^{i}}\right|
_{\Sigma }$ \emph{has\textbf{\ maximal rank},} $\Re (\mathbf{J})=R.$

A simple classical mechanical example of functionally \emph{dependent}
constraints occurs in a $2$-dimensional phase space with coordinates $\left(
q,p\right) $ and constraints $\phi ^{1}\equiv q\approx 0$ and $\phi
^{2}\equiv pq\approx 0$. In this case, $\mathbf{J}=\left[ 
\begin{array}{ll}
1 & p \\ 
0 & q
\end{array}
\right] _{q=0}$ and $\Re \left( \mathbf{J}\right) =1$.

A system of just one constraint can also fail the test of regularity.
Consider for example the constraint $\phi =q^{2}\approx 0$ in a $2$%
-dimensional phase space. In this case, $\mathbf{J}=\left( 2q,0\right)
_{q^{2}=0}=0$ and $\Re \left( \mathbf{J}\right) =0$. The same problem occurs
with the constraint $q^{k}\approx 0$, for $k>1$, which has a zero of $k$-th
order on the constraint surface. This example illustrates that one \emph{%
constraint} may be dependent on itself, while one \emph{function} is, by
definition, always functionally independent.\medskip

\textbf{Equivalence}: \emph{Different sets of constraints are said to be
equivalent if they define the same constraint surface.}\medskip

Note that this definition refers to the locus of constraints, not to
equivalence of the resulting dynamics. Since the surface $\Sigma $ is
defined by the zeros of the constraints, while the regularity conditions
depend on their derivatives, it is possible to replace a set of irregular $%
\phi $'s by an \emph{equivalent} set of \emph{regular} constraints $\tilde{%
\phi}$.

In the classification of irregular systems, two questions present
themselves: what is the \emph{nature} of the constraints that give rise to
irregularity, and \emph{where} the irregularities can occur. These issues
are addressed in the following subsections. A third question is whether a
system can evolve from an initial state in which regularity holds, into an
irregular configuration. This will be discussed in the last section.

\subsection{Basic types of irregular constraints}

Irregular constraints can be classified according to their behavior in the
vicinity of the surface $\Sigma $. For example, linearly dependent
constraints have Jacobian with constant rank $R^{\prime }$ throughout $%
\Sigma $, and 
\begin{equation}
\phi ^{r}\equiv J_{i}^{r}(\bar{z})\left( z^{i}-\bar{z}^{i}\right) \approx
0,\qquad \Re (\mathbf{J})=R^{\prime }<R.  \label{linear}
\end{equation}
These constraints are regular systems in disguise simply because $%
R-R^{\prime }$ constraints are redundant and should be discarded. The subset
with $R^{\prime }$ linearly independent constraints gives the correct
description. For example, the linearly dependent constraints $\phi ^{1}=z$
and $\phi ^{2}=2z$ are clearly in this category.\ Apart from this trivial
case, two main types of truly irregular constraints, which do not possess a
linear approximation in the vicinity of some points of $\Sigma $\textbf{,}
can be distinguished:

\subparagraph{\textbf{Type I. Multilinear constraints}.}

Consider the constraint 
\begin{equation}
\phi \equiv \prod\limits_{i=1}^{M}f_{i}(z)\approx 0,  \label{multilinear}
\end{equation}
where the functions $f_{i}$ have simple zeros. Each factor defines a surface
of codimension 1, 
\begin{equation}
\Sigma _{i}\equiv \left\{ \bar{z}\in \Gamma \mid f_{i}(\bar{z})=0\right\} ,
\label{codimension 1}
\end{equation}
and $\Sigma $ is the collection of all surfaces, $\Sigma =\bigcup \Sigma
_{i}.$ The rank of Jacobian of $\phi $ is reduced at intersections 
\begin{equation}
\Sigma _{ij}\equiv \Sigma _{i}\bigcap \Sigma _{j}.  \label{intersections}
\end{equation}
Thus, the RCs hold everywhere on $\Sigma $, except at the intersections $%
\Sigma _{ij}$, where $\phi $ has zeros of higher order. Note that the
intersections (\ref{intersections}) also include the points where more than
two $\Sigma $'s overlap.

\subparagraph{\textbf{Type II. Nonlinear constraints.}}

Consider the constraint of the form 
\begin{equation}
\phi \equiv \left[ f(z)\right] ^{k}\approx 0\qquad \left( k>1\right) ,
\label{nonlinear}
\end{equation}
where the function $f(z)$\ has a simple zero. This constraint has a zero of
order $k$ in the vicinity of $\Sigma $, its Jacobian vanishes on the
constraint surface and, therefore, the RCs fail (here we assume $k>1$ in
order to avoid infinite values for $\frac{\partial \phi }{\partial z^{i}}$
on $\Sigma $). It could seem harmless to replace $\phi $ by the equivalent
regular constraint $f(z)\approx 0$, but it turns out that this may change
the dynamics of original system, as we show below.\medskip

Types I and II are the two fundamental generic classes of irregular
constraints. In general, there can be combinations of them occurring
simultaneously near a constraint surface, as in constraints of the form $%
\phi =\left[ f_{1}(z)\right] ^{2}f_{2}(z)\approx 0$, etc.

\subsection{Classification of constraint surfaces}

The previous classification refers to the way in which $\phi $ approaches
zero. Now we will discuss \emph{where} regularity can be violated. The rank
of the Jacobian $\frac{\partial \phi ^{r}}{\partial z^{i}}$ need not be
constant throughout $\Sigma :$ suppose one eigenvalue of the Jacobian
vanishes on a submanifold $\Sigma _{0}\subset \Sigma $. On $\Sigma _{0}$
regularity is violated, while it still holds on the rest of $\Sigma $. Thus,
barring accidental degeneracies such as linearly dependent constraints, one
of these three situation may present themselves:\medskip

\textbf{A}. \emph{The RCs are satisfied everywhere on the constraint surface}%
: $\mathbf{J}$ has maximal rank throughout $\Sigma $ (regular systems).

\textbf{B}. \emph{The RCs fail everywhere on the constraint surface:} $%
\mathbf{J}$ has constant rank $R^{\prime }<R$ on $\Sigma .$

\textbf{C}. \emph{The RCs fail on $\Sigma _{0}$: }$\left. \Re \left( \mathbf{%
J}\right) \right| _{\Sigma _{0}}=R^{\prime }<R$, while $\Re \left( \mathbf{J}%
\right) =R$ elsewhere on $\Sigma $.\medskip

In the last case, the constraint surface can be decomposed into two non
overlapping sets $\Sigma _{0}$ and $\Sigma _{R}$. Then, the rank of the
Jacobian jumps from $\Re \left( \mathbf{J}\right) =R$ on $\Sigma _{R}$, to $%
\Re \left( \mathbf{J}\right) =R^{\prime }$ on $\Sigma _{0}$. Although the
functions $\phi ^{r}$ are continuous and differentiable, this is not
sufficient for regularity. Irregular cases are illustrated by the following
examples.

In a $(2+N)$-dimensional phase space $(q,p,z^{1},\ldots ,z^{N})$, the
constraints $\phi ^{1}\equiv q-p\approx 0$\ and $\phi ^{2}\equiv qp\approx 0$
are irregular on the whole constraint surface $\left\{ (0,0,z^{1},\ldots
,z^{N})\right\} ,$ where the Jacobian has rank $\Re (\mathbf{J})=1$. Note
that these constraints are always irregular, although the functions $q+p$
and $qp$ are functionally independent everywhere except for $q=p$, which
happens to be the case at the constraint surface.

An example having both regular and irregular sectors is a massless
relativistic particle\ in Minkowski space with phase space $(q^{\mu },p_{\nu
})$. The constraint $\phi \equiv p^{\mu }p_{\mu }\approx 0\;$has Jacobian $%
\mathbf{J}=(0,2p^{\mu })_{\phi =0}$, and its rank is one everywhere, except
at the apex of the cone, $p^{\mu }=0$, where the light-cone is not
differentiable and the Jacobian has rank zero.

The lack of regularity, however, is not necessarily due to the absence of a
well defined smooth tangent space for $\Sigma $. Consider for example the
multilinear constraint 
\begin{equation}
\phi (x,y,z)=(x-1)(x^{2}+y^{2}-1)\approx 0.
\end{equation}
Here the constraint surface $\Sigma $ is composed of two sub-manifolds: the
plane $\Pi =\{(x,y,z)\mid $ $x-1\approx 0\}$, and the cylinder $%
C=\{(x,y,z)\mid $ $x^{2}+y^{2}-1\approx 0\}$, which are tangent to each
other along the line $L=\{(x,y,z)\mid $ $x=1,y=0,z\in \Bbb{R}$. The Jacobian
on $\Sigma $ is 
\begin{equation}
\mathbf{J}=\left( 3x^{2}+y^{2}-2x-1,2y\left( x-1\right) ,0\right) _{\phi =0}
\end{equation}
and its rank is $1$ everywhere, except on $L$, where it is zero. The
constraint $\phi $ is irregular on this line. However, the tangent vectors
to $\Sigma $ are well defined there. The irregularity arises because $\phi $
is a multilinear constraint of the type described by (\ref{multilinear}) and
has two simple zeros overlapping on $L$. The equivalent set of regular
constraints on $L$ is $\{\phi _{\Pi }=x-1\approx 0,\,\phi
_{C}=x^{2}+y^{2}-1\approx 0\}$, as we will see below.

\section{Treatment of irregular systems}

In what follows regular systems and linearly dependent constraints will not
be discussed. They are either treated in standard texts, or they can be
trivially reduced to the regular case.

\subsection{Multilinear constraints}

Consider a system of type I, as in Eq. (\ref{multilinear}). In the vicinity
of an irregular point where only two surfaces (\ref{codimension 1})
intersect, say $\Sigma _{1}$ and $\Sigma _{2}$, the constraint $\phi \approx
0$ is equivalently described by the set of regular constraints

\begin{equation}
f_{1}\approx 0,\qquad f_{2}\approx 0.  \label{regular set}
\end{equation}
This replacement generically changes the Lagrangian of the system, and the
orbits, as well. Suppose the original canonical Lagrangian is

\begin{equation}
L(q,u)=p_{i}\dot{q}^{i}-H(q,p)-u\phi (q,p),  \label{LC old}
\end{equation}
where $H$ is the Hamiltonian containing all regular constraints. Replacing $%
\phi $ by (\ref{regular set}), gives rise to an effective Lagrangian 
\begin{equation}
L_{12}(q,v)=p_{i}\dot{q}^{i}-H(q,p)-v^{1}f_{1}(q,p)-v^{2}f_{2}(q,p).
\label{LC new}
\end{equation}
defined on $\Sigma _{12}$. Thus, instead of the \emph{irregular} Lagrangian (%
\ref{LC old}) defined on the whole $\Sigma $, there is a collection of \emph{%
regularized }effective Lagrangians defined in the neighborhood of the
different intersections of $\Sigma _{i}$s. For each of these regularized
Lagrangians, the Dirac procedure can be carried out to the end.

Let us illustrate this with the example of a Lagrangian in a $\left(
2+N\right) $-dimensional configuration space $(x,y,q^{1},\ldots ,q^{N})$, 
\begin{equation}
L=\frac{1}{2}\,\sum\limits_{k=1}^{N}\left( \dot{q}^{k}\right) ^{2}+\frac{1}{2%
}\,\left( \dot{x}^{2}+\dot{y}^{2}\right) -\lambda xy.  \label{example}
\end{equation}
This Lagrangian describes a free particle moving\ on the set 
\begin{equation}
\left\{ \left. \left( x,y,q^{k}\right) \in \Bbb{R}^{N+2}\right| xy=0\right\}
\subset \Bbb{R}^{N+2},
\end{equation}
which is the union of two $(N+1)$-dimensional planes where $x$ and $y$
vanish respectively. The constraint surface defined by $xy=0$ can be divided
into the following sets: 
\begin{eqnarray}
\Sigma _{1} &=&\left\{ \left. \left( x,0,q^{k};p_{x},p_{y},p_{k}\right)
\right| x\neq 0\right\}   \nonumber \\
\Sigma _{2} &=&\left\{ \left. \left( 0,y,q^{k};p_{x},p_{y},p_{k}\right)
\right| y\neq 0\right\}   \label{sigma-y} \\
\Sigma _{12} &=&\left\{ \left( 0,0,q^{k};p_{x},p_{y},p_{k}\right) \right\} .
\nonumber
\end{eqnarray}
The constraint is regular on $\Sigma _{1}\bigcup \Sigma _{2}$, while on $%
\Sigma _{12}$ it is irregular and can be exchanged by $\left\{ \phi
_{1}=x\approx 0\right. $, $\left. \phi _{2}=y\approx 0\right\} $. The
corresponding regularized Lagrangians are 
\begin{eqnarray}
L_{1} &=&\frac{1}{2}\,\sum\limits_{k=1}^{N}\left( \dot{q}^{k}\right) ^{2}+%
\frac{1}{2}\,\dot{x}^{2},  \nonumber \\
L_{2} &=&\frac{1}{2}\,\sum\limits_{k=1}^{N}\left( \dot{q}^{k}\right) ^{2}+%
\frac{1}{2}\,\dot{y}^{2}  \label{L2} \\
L_{12} &=&\frac{1}{2}\,\sum\limits_{k=1}^{N}\left( \dot{q}^{k}\right) ^{2}, 
\nonumber
\end{eqnarray}
and the Lagrange multipliers have dropped out, so the regularized
Lagrangians describe physical degrees of freedom only -- as expected.

The corresponding regularized Hamiltonians are 
\begin{eqnarray}
H_{1} &=&\frac{1}{2}\,\sum\limits_{k=1}^{N}p_{k}^{2}+\frac{1}{2}\,p_{x}^{2},
\nonumber \\
H_{2} &=&\frac{1}{2}\,\sum\limits_{k=1}^{N}p_{k}^{2}+\frac{1}{2}\,p_{y}^{2}
\label{H2} \\
H_{12} &=&\frac{1}{2}\,\sum\limits_{k=1}^{N}p_{k}^{2},  \nonumber
\end{eqnarray}
which are defined in the corresponding reduced manifolds of phase space
(obtained after completing the Dirac procedure): 
\begin{eqnarray}
\tilde{\Sigma}_{1} &=&\left\{ \left. \left( x,0,q^{k};p_{x},0,p_{k}\right)
\right| x\neq 0\right\}  \nonumber \\
\tilde{\Sigma}_{2} &=&\left\{ \left. \left( 0,y,q^{k};0,p_{y},p_{k}\right)
\right| y\neq 0\right\}  \label{sigma total} \\
\tilde{\Sigma}_{12} &=&\left\{ \left( 0,0,q^{k};0,0,p_{k}\right) \right\} . 
\nonumber
\end{eqnarray}

It is straightforward to generalize the proposed treatment when more than
two surfaces $\Sigma _{i}$ overlap.

\subparagraph{Evolution of a multilinearly constrained system.}

Since in the presence of a multilinear constraint there are regions of the
phase space where the Jacobian has different rank, a question arises about
the evolution of this system. Can the system evolve from a generic
configuration in a region of maximal rank, reaching a configuration of lower
rank in finite time? In the case that that were possible, what happens with
the system afterwards?. (This problem should not be confused with the issues
arising in degenerate systems \cite{Degenerate}-\cite{Generic}.)

To answer this question let us consider the simple example discussed above (%
\ref{example}), for $N=1$, 
\begin{equation}
L=\frac{1}{2}\,\left( \dot{x}^{2}+\dot{y}^{2}+\dot{z}^{2}\right) -\lambda xy.
\label{N=1}
\end{equation}
In the regions $\Sigma _{1}$ and $\Sigma _{2}$ [see Eqs. (\ref{sigma-y})],
the rank is maximal and the free particle can move freely along the $x$- or $%
y$-axis respectively.

Suppose that the initial state is 
\begin{equation}
x(0)=a>0,\quad y(0)=0,\quad z(0)=0,\quad \dot{x}(0)=-v<0,\quad \dot{y}%
(0)=0,\quad \dot{z}(0)=0,  \label{t=0}
\end{equation}
so that the particle is moving on $\Sigma _{1}$, with finite speed along the 
$x$-axis towards $x=0$ on $\Sigma _{12}$. The evolution is given by $\bar{x}%
(t)=a-vt$, $\bar{y}(t)=0$, $\bar{z}(t)=0$ and the particle clearly reaches $%
x=0$ in a finite time ($T=a/v$). What happens then? According to the
evolution equation, for $x<0$ the trajectory takes the form $\bar{x}%
(t)=a^{\prime }-v^{\prime }t$, $\bar{y}(t)=0$, $\bar{z}(t)=0$, however the
action would be infinite unless $a=a^{\prime }$ and $v=v^{\prime }$.
Therefore, the particle continues unperturbed past beyond the point where
the RCs fail. So, the irregular surface is not only reachable in a finite
time, but it is crossed without any observable effect on the trajectory.

>From the point of view of the trajectory in phase space, it is clear that
the initial state $\left( a,0,0;-v,0,0\right) $ lies on the surface $\tilde{%
\Sigma}_{1}$, and at $t=T$ the system reaches the point $\left(
0,0,0;-v,0,0\right) $, which \emph{does not lie on} the surface $\tilde{%
\Sigma}_{12}=\left\{ \left( 0,0,z;0,0,p_{z}\right) \right\} $.

While it is true that at $t=T$ the Jacobian changes rank, it would be
incorrect to conclude that the evolution suffers a jump since the dynamical
equations are perfectly valid there. In order to have significant change in
dynamics, the Jacobian should change its rank in an open set.

\subsection{Nonlinear constraints}

Let us now turn to the case of irregular systems of type II. As we will
show, it is possible to replace a nonlinear irregular constraint by an
equivalent linear one without changing the dynamical contents of the theory,
provided the linear constraint is second class. Otherwise, the resulting
Hamiltonian dynamics will be, in general, inequivalent to that of the
original Lagrangian system.

In order to illustrate this point, consider a system given by the Lagrangian 
\begin{equation}
L(q,u)=\frac{1}{2}\,\gamma _{ij}\,\dot{q}^{i}\dot{q}^{j}-u\left[ f(q)\right]
^{k},  \label{ex-1}
\end{equation}
where $k>1$\ and

\begin{equation}
f(q)\equiv c_{i}q^{i}\neq 0,\quad i=1,...,N.
\end{equation}
Here we assume the metric $\gamma _{ij}$ to be constant and invertible, and
the coefficients $c_{i}$ are also constant. The Euler-Lagrange equations
describe a free particle in an $N$-dimensional space, with time evolution $%
\bar{q}^{i}(t)=v_{0}^{i}\,t+q_{0}^{i}$, where $u(t)$ is a Lagrange
multiplier. This solution is determined by $2N$ initial conditions, $%
q^{i}(0)=q_{0}^{i}$ and $\dot{q}^{i}(0)=v_{0}^{i}$ subject to the
constraints $c_{i}q_{0}^{i}=0$ and $c_{i}v_{0}^{i}=0$. Thus, the system
possesses $N-1$ physical degrees of freedom.

In the Hamiltonian approach this system has a primary constraint $\pi \equiv 
\frac{\partial L_{2}}{\partial \dot{u}}\approx 0$\ whose preservation in
time leads to the secondary constraint 
\begin{equation}
\phi \equiv \left[ f(q)\right] ^{k}\approx 0.  \label{phi}
\end{equation}
According to (\ref{nonlinear}), this is a nonlinear constraint and there are
no further constraints. As a consequence, the system has only two first
class constraints $\left\{ \pi \approx 0,\;f^{k}\approx 0\right\} $, and $%
N-1 $ degrees of freedom, as found in the Lagrangian approach.

On the other hand, if one chooses instead of (\ref{phi}), the equivalent
linear constraint 
\begin{equation}
f(q)=c_{i}q^{i}\approx 0,  \label{root}
\end{equation}
then its time evolution yields a \emph{new} constraint, 
\begin{equation}
\chi (p)\equiv \gamma ^{ij}c_{i}\,p_{j}\approx 0.
\end{equation}
Now, since 
\begin{equation}
\left\{ f,\chi \right\} =\gamma ^{ij}c_{i}\,c_{j}\equiv \left\| c\right\|
^{2},
\end{equation}
two cases can be distinguished:

\begin{itemize}
\item  If $\left\| c\right\| =0$, there are three first class constraints, $%
\pi \approx 0$, $f\approx 0$ and $\chi \approx 0$, which means that the
system has $N-2$\ physical degrees of freedom. In this case, substitution of
(\ref{phi}) by the equivalent linear constraint (\ref{root}), yields a \emph{%
dynamically inequivalent} system.

\item  If $\left\| c\right\| \neq 0$, then $f\approx 0$ and $\chi \approx 0$
are second class, while $\pi \approx 0$ is first class, which leaves $N-1$
physical degrees of freedom\ and the substitution does not change the
dynamics of the system.
\end{itemize}

Thus, if $f^{k}\approx 0$ is irregular, replacing it by the regular
constraint $f\approx 0$ changes the dynamics if $f$ is a first class
function, but it gives the correct result if it is a second class function.

Note that in the Lagrangian description there is no room to distinguish
first and second class constraints, so it would seem like the value of $%
\mathbf{||}c||$ didn't matter. However, the inequivalence of the
substitution can be understood in the Lagrangian analysis as well. Suppose
that it were permissible to exchange the constraint $f^{k}\approx 0$ by $%
f\approx 0$ in the Lagrangian. Then, instead of (\ref{ex-1}), one would have 
\begin{equation}
\tilde{L}(q,u)=\frac{1}{2}\,\gamma _{ij}\,\dot{q}^{i}\dot{q}^{j}-uf(q).
\label{ex-2}
\end{equation}
It can be easily checked that (\ref{ex-2}) yields $N-2$ degrees of freedom
when $\left\| c\right\| =0$, and $N-1$ degrees of freedom when $\left\|
c\right\| \neq 0$, which agrees with the results obtained in the Hamiltonian
analysis. Note that the substitution of $f^{k}$ by $f$ modifies the dynamics
only if $\gamma ^{ij}c_{i}\,c_{j}=0$, but this can happen nontrivially only
if the metric $\gamma _{ij}$ is not positive definite.

In general, a nonlinear irregular constraint $\phi \approx 0$\ has a
multiple zero on the constraint surface $\Sigma $, which means that its
gradient vanishes on $\Sigma $ as well. An immediate consequence of $\left(
\partial \phi /\partial z^{i}\right) \approx $ $0$, is that $\phi $ commutes
with all \emph{finite} functions on $\Gamma $, 
\begin{equation}
\left\{ \phi ,F(z)\right\} \approx 0.  \label{pathological 1}
\end{equation}
As a consequence, $\phi \approx 0$ is first class and is always preserved in
time, 
\begin{equation}
\dot{\phi}\approx 0.
\end{equation}
On the other hand, a nonlinear constraint cannot be viewed as a symmetry
generator simply because it does not generate any transformation, 
\begin{equation}
\delta _{\varepsilon }z^{i}=\left\{ z^{i},\varepsilon \phi \right\} \approx
0.  \label{trivial}
\end{equation}
Consistently with this, $\phi $ cannot be gauge-fixed, as there is no finite
function $\mathcal{P}$ on $\Gamma $ such that 
\begin{equation}
\left\{ \phi ,\mathcal{P}\right\} \neq 0.  \label{pathological 2}
\end{equation}

In this sense, a nonlinear first class constraint that cannot be
gauge-fixed, cancels only half a degree of freedom. The other half degree of
freedom cannot be cancelled because the gauge-fixing function does not exist
and, in particular, it cannot appear in the Hamiltonian. Although the
features (\ref{pathological 1}-\ref{pathological 2}) allow counting the
degrees of freedom in a theory, these systems are pathological and their
physical relevance is questionable since their Lagrangians cannot be
regularized.

When a nonlinear constraint $\phi \approx 0$ can be exchanged by a regular
one, the Lagrangian is regularized as in the case of multilinear
constraints. For example, the system (\ref{ex-1}) with $\left\| c\right\|
\neq 0$ has Hamiltonian 
\begin{equation}
H=\frac{1}{2}\,\gamma ^{ij}p_{i}p_{j}+\lambda \pi +uf(q),
\end{equation}
where $f=c_{i}q^{i}$ will turn out to be a second class constraint. The
corresponding regularized Lagrangian is 
\begin{equation}
L_{reg}=\frac{1}{2}\,\gamma _{ij}\,\dot{q}^{i}\dot{q}^{j}-uf(q),
\end{equation}
which coincides with $\tilde{L}$, Eq. (\ref{ex-2}), as expected.

In the Refs. \cite{Garcia-Pons,Pons-Salisbury-Shepley} irregular systems of
the type II were discussed. It was pointed out that there was a possible
loss of dynamical information in some cases. From our point of view, it is
clear that this would occur when $f$ is a first class function.

\section{\textbf{\textrm{Linearization of irregular systems}}}

It has been observed in five dimensional Chern Simons theory, that\textbf{\ }%
the effective action for the linearized perturbations of the system around
certain backgrounds seems to have more degrees of freedom than the fully
nonlinear theory \cite{Chandia-Troncoso-Zanelli}. This is puzzling since the
heuristic picture is that the degrees of freedom of a system correspond to
the small perturbations around a local minimum of the action, and therefore
the number of degrees of freedom should not change when the linearized
approximation is used.

In view of the discussion in the previous section, it is clear that a
possible solution of the puzzle lies in the fact that substituting a
nonlinear constraint by a linear ones may change the dynamical features of
the theory. But the problem with linear approximations is more serious: the
linearized approximation retains only up to quadratic and bilinear terms in
the Lagrangian, which give linear equations for the perturbations. Thus,
irregular constraints in the vicinity of the constraint surface are erased
in the linearized action. The smaller number of constraints in the effective
theory can lead to the wrong conclusion that the effective system possess
more degrees of freedom than the unperturbed theory. The lesson to be
learned is that the linear approximation is not valid in the part of the
phase space where the RCs fail.

This is illustrated by the same example discussed earlier (\ref{ex-1}). One
can choose as a background the solution $(\bar{q}^{1},\ldots ,\bar{q}^{N},%
\bar{u})$, where $\bar{q}^{i}(t)=q_{0}^{i}+v_{0}^{i}t$ satisfies the
constraint 
\begin{equation}
c_{i}\bar{q}^{i}=0,  \label{fq}
\end{equation}
and $\bar{u}(t)$ is an arbitrarily given function. This describes a free
particle moving in the ($N-1$)-dimensional plane defined by (\ref{fq}). The
linearized effective Lagrangian, to second order in the small perturbations $%
s^{i}=q^{i}-\bar{q}^{i}(t)$ and $w=u-\bar{u}(t)$, has the form 
\begin{equation}
L_{e\!f\!f}\left( s,w\right) =\frac{1}{2}\,\gamma _{ij}\left( v_{0}^{i}+\dot{%
s}^{i}\right) \left( v_{0}^{j}+\dot{s}^{j}\right) -\bar{u}\left(
c_{i}s^{i}\right) ^{2},  \label{Leff}
\end{equation}
and the equations of motion are 
\begin{equation}
\ddot{s}^{i}+\Gamma _{j}^{i}(t)\,s^{j}=0\qquad i=1,\ldots ,N,  \label{Lineq}
\end{equation}
where $\Gamma _{j}^{i}\equiv 2\bar{u}\,\gamma ^{ik}c_{k}c_{j}$ is the eigen
frequency matrix. Since $\bar{u}$\ is not a dynamical variable, it is not
varied and the nonlinear constraint $\left( c_{i}s^{i}\right) ^{2}=0$ is
absent from the linearized equations. The system described by (\ref{Lineq})
possesses $N$ physical degrees of freedom, that is, one degree of freedom
more than the original nonlinear theory (\ref{ex-1}).

The only indication that one of these degrees of freedom has a nonphysical
origin is the following: If $\Vert c\Vert \neq 0$, splitting the components
of $s^{i}$ along $c_{i}$ and orthogonal to $c_{i}$\ as 
\begin{equation}
s^{i}(t)\equiv s(t)\gamma ^{ij}c_{j}+s_{\perp }^{i}(t),
\end{equation}
gives rise to the projected equations 
\begin{eqnarray}
\ddot{s}_{\perp }^{i} &=&0  \label{s-perp} \\
\ddot{s}+2\bar{u}(t)\Vert c\Vert ^{2}s &=&0.  \label{s}
\end{eqnarray}

The $N-1$ components of $s_{\perp }^{i}(t)$ obey a deterministic second
order equation, whereas $s(t)$ satisfies an equation which depends on an
indeterminate arbitrary function $\bar{u}(t)$. The dependence of $s=\bar{s}%
(t,\bar{u}(t))$ on the background Lagrange multiplier $\bar{u}$ is an
indication that $s$ is a nonphysical degree of freedom, since $u$ was an
arbitrary function to begin with. This is not manifest in Eq. (\ref{s}),
where $\bar{u}$\ is a fixed function and, from a naive point of view, $s(t)$
is determined by the same equation, regardless of the physically obscure
origin of the function $\bar{u}$. It is this naive analysis that leads to
the wrong conclusion indicated above.

Let us emphasize that a linearized theory may be consistent by itself, but
it is not necessarily a faithful approximation of a nonlinear theory.

One way to avoid the inconsistencies between the original theory and the
linearized one would be to first regularize the constraints (if possible)
and then linearize the corresponding regular Lagrangian.

\section{Chern-Simons theories}

Hamiltonian structure of CS theories has been studied in \cite
{Banados-Garay-Henneaux}. The phase space of a CS theory in $D=2n+1$
space-time dimensions, invariant under $N$-parameter gauge group, is defined
by canonically conjugate pairs of fields $(A_{i}^{a}(x,t),\pi
_{a}^{i}(x^{\prime },t))$, where $a=1,\ldots ,N$ and $x^{i}$ ($i=1,\ldots
,2n $) are the local coordinates on a spatial section. The CS Hamiltonian
density is given by 
\begin{equation}
\mathcal{H}=A_{0}^{a}\,\tilde{G}_{a}+u_{i}^{a}\phi _{a}^{i},
\end{equation}
where $u_{i}^{a}(x,t)$ and $A_{0}^{a}(x,t)$ are Lagrange multipliers for the
constraints 
\begin{eqnarray}
\phi _{a}^{i} &\equiv &\pi _{a}^{i}-\mathcal{L}_{a}^{i}(A_{j}^{b})\approx 0,
\label{primary} \\
G_{a} &=&g_{aa_{1}\cdots a_{n}}F^{a_{1}}\wedge \cdots \wedge
F^{a_{n}}\approx 0.  \label{G}
\end{eqnarray}
Here $G_{a}\equiv d^{2n}x\,\tilde{G}_{a}$, $g$ is a symmetric tensor of rank 
$n+1$, invariant under action of a gauge group, and $F=dA+A\wedge A$ is the
curvature 2-form associated to the gauge field 1-form $A$.

Constraints $\phi _{a}^{i}$ are regular because they are linear in momenta.
Thus, the regularity of CS theories is determined by momentum-independent
constraints $G_{a}$. Their small variations, $\delta G_{a}=\mathbf{J}%
_{ab}\,D\delta A^{b}$, evaluated at $G_{a}=0$, give the $(2n-2)$-form $%
\mathbf{J}_{ab}$, which can be identified as the Jacobian, 
\begin{equation}
\mathbf{J}_{ab}\equiv n\,g_{aba_{2}\cdots a_{n}}\,F^{a_{2}}\wedge \cdots
\wedge F^{a_{n}}.  \label{CS Jacobian}
\end{equation}
According to Dirac's definition, sufficient and necessary condition for $%
G_{a}$ to be regular is 
\begin{equation}
\Re \left( \mathbf{J}_{ab}\right) =N.  \label{rc}
\end{equation}
Since $\mathbf{J}_{ab}$ is field dependent, its rank may change in space. In
particular, for a pure gauge configuration $F=0$, and $\mathbf{J}_{ab}$ has
rank zero. For other configurations, the rank of Jacobian can range from
zero to $N$, and the irregularities are always of multilinear type because
in the expression (\ref{G}) the phase space coordinates $A_{i}^{a}$ occur
only linearly.

In general, a non-abelian CS theory for $D\geq 5$ possesses a non-vanishing
number of physical degrees of freedom \cite{Banados-Garay-Henneaux} 
\begin{equation}
f_{2n+1}(N)=nN-n-N\qquad (N\geq 2)  \label{generic degrees}
\end{equation}
in regular and \emph{generic }cases \cite{Generic}. Therefore, in CS
theories, the study of dynamics requires not only the analysis of
regularity, but also of genericity. In spite of the fact that both
conditions are expressed in terms of the same matrix $\Omega _{ab}^{ij}$,
they are independent. For example, the extreme case of $F=0$ is both
irregular and degenerate, but there are examples in CS supergravity which
are generic and irregular \cite{Chandia-Troncoso-Zanelli}. The opposite case
occurs in a five-dimensional CS theory based on $G_{1}\times G_{2}$ for
particular choice of invariant tensor. In this case, there exist
configurations which are regular but degenerate. Take the group indices as $%
a=(r,\alpha )$ corresponding to $G_{1}$ and $G_{2}$ respectively, and
invariant tensor as $g_{rs1}=g_{rs}$ and $g_{\alpha \beta \bar{1}}=g_{\alpha
\beta }$ (both invertible). Then the configuration $F^{a}=\left(
f^{1}dx^{1}\wedge dx^{2},h^{\bar{1}}dx^{3}\wedge dx^{4}\right) $ is regular
and degenerate. Indeed, $\mathbf{J}_{ab}=\left( 
\begin{array}{cc}
g_{rs}\,f^{1} & 0 \\ 
0 & g_{\alpha \beta }\,h^{\bar{1}}
\end{array}
\right) $ is regular, while $\Omega _{ab}^{ij}$ with non-vanishing
components $\Omega _{rs}^{34}=g_{rs}\,f^{1}$ and $\Omega _{\alpha \beta
}^{12}=g_{\alpha \beta }\,h^{\bar{1}}$ has $2N$ zero modes and is therefore
degenerate.

As a consequence of existence of both regularity and genericity issues, the
regularization problem is much more delicate in CS theories.

\section{\textbf{\textrm{Comments}}}

\subparagraph{1 - Dirac conjecture.}

Dirac conjectured that \emph{all} first class constraints generate gauge
symmetries \cite{DiracCJM}. It was shown that Dirac's conjecture is not true
for first class constraints of the form $f^{k}$ $(k>1)$, and following from $%
\dot{f}\approx 0$ \cite{Castellani,Blagojevic}. Therefore, for systems with
nonlinear constraints, the conjecture does not work and they generically
provide counterexamples of it \cite{Henneaux-Teitelboim,Allcock,Cawley}.

From the point of view of irregular systems, it is clear that Dirac's
conjecture fails for nonlinear constraints because they do not generate any
local transformation, c.f. Eq. (\ref{trivial}). In Refs. \cite
{Garcia-Pons,Pons-Salisbury-Shepley} it was observed that Dirac's conjecture
may not hold in the presence of irregular constraints of type II.

In the case of multilinear constraints, however, Dirac's conjecture holds.
The fact that at irregular points the constraints do not generate any
transformation only means that these are fixed points of the gauge
transformation.

\subparagraph{2 - Quantization.}

Although, in view of the above discussion, it is possible to deal
systematically with classical theories containing irregular constraints,
there may be severe problems in their quantum description. Consider a path
integral of the form 
\begin{equation}
Z\sim \int [dq][dp][du]\exp i\left[ p\dot{q}-H(q,p)-u\phi (q,p)\right] ,
\end{equation}
where $\phi =\left[ f(q,p)\right] ^{k}$ is a nonlinear constraint.
Integration on $u$ yields to $\delta \left( f^{k}\right) $, which is not
well-defined for a zero of order $k>1$, making the quantum theory ill
defined. Only if the nonlinear constraint could be exchanged by the regular
one, $f(q,p)\approx 0$, the quantum theory could be saved. An example of
this occurs in the standard approach for QED, where it is usual practice to
introduce the nonlinear gauge fixing term $u(\partial _{i}A^{i})^{2}$ in
order to fix the primary constraint $\pi ^{0}\equiv \left( \delta
I_{ED}/\delta \dot{A}_{0}\right) \approx 0$. Since the gauge condition $%
f(A)=\partial _{i}A^{i}(x)\approx 0$ is a second class constraint, its
substitution by a regular constraint does not change its dynamical structure.

\section{\textbf{\textrm{Summary}}}

We have discussed the dynamics and evolution of a system possessing
constraints which may violate the regularity conditions (functional
independence) on some subsets of the constraint surface $\Sigma $. These
so-called irregular systems are seen to arise generically because of
nonlinearities in the constraints and can be classified into two families:
multilinear (type I) and nonlinear (type II).

\begin{itemize}
\item  Type I constraints are of the form $\phi =\prod f_{i}(z)$, where $%
f_{i}$ possess simple zeros. These constraints violate the regularity
conditions (RCs) on sets of measure zero on the constraint surface $\Sigma $.

\item  Type I constraints can be exchanged by equivalent constraints which
are regular giving an equivalent dynamical system.

\item  Type II constraints are of the form $\phi =f^{k}$ $(k>1)$ where $f$
has a simple zero. They violate the RCs on sets of nonzero measure on $%
\Sigma $.

\item  A type II constraint can be replaced by an equivalent linear one only
if the latter is second class; if the equivalent linear constraint is first
class, substituting it for the original constraint would change the system.

\item  In general, the orbits can cross the configurations where the RCs are
violated without any catastrophic effect for the system. If the symplectic
form degenerates at the irregular points, additional analysis is required.

\item  The naive linearized approximation of an irregular constrained system
generically changes it by erasing the irregular constraints. In order to
study the perturbations around a classical orbit in an irregular system, it
would be necessary to first regularize it (if possible) and only then do the
linearized approximation.

\item  Chern-Simons theories possess irregular and regular sectors. This
problem is independent of the presence of degeneracies in the symplectic
form, making the regularization problem much more complex.
\end{itemize}

\section{\textbf{\textrm{Acknowledgments}}}

We are grateful to Milutin Blagojevi\'{c}, Marc Henneaux and Claudio
Teitelboim for enlightening and helpful comments, and specially to Ricardo
Troncoso for many useful discussions, insights and friendly criticisms. We
would also like to thank Josep Mar\'{i}a Pons and Antonio Garc\'{i}a-Zenteno
for bringing their work to our attention and for indicating its relevance to
ours. This work is partially funded by grants FONDECYT 1020629, 7020629,
1010450, 2010017 and the grant MECESUP USA 9930\textbf{. }One of us (O. M.)
thanks the Abdus Salam ICTP for hospitality during the completion of this
work. The generous support of Empresas CMPC to CECS is also acknowledged.
CECS is a Millennium Science Institute and is funded in part by grants from
Fundaci\'{o}n Andes and the Tinker Foundation.


\begin{thebibliography}{99}
\bibitem{Degenerate}  These questions have a similar ring to those presented
in the case of a degenerate dynamical system, in which the symplectic matrix
changes its rank \cite{Saavedra-Troncoso-Zanelli}: Is it possible for a
system in a generic initial configuration, to reach (in a finite time) a
configuration where the symplectic form has lower rank? That problem is
unrelated to the one discussed here and can be treated independently. In the
present case it is the functional independence of the constraints that
fails; in the degenerate dynamical systems it is the symplectic structure
that breaks down.
\bibitem{Generic}  A configuration is called generic if the symplectic form $%
\Omega $ is invertible. Otherwise, it is called degenerate. In the case of
CS theories, the symplectic form is given by $d^{2n}x\,\Omega
_{ab}^{ij}\equiv dx^{i}\wedge dx^{j}\wedge J_{ab}$. In these theories, the
generic configurations are those for which (\emph{i}) the $2n$ independent
vectors $\left( \mathbf{V}_{i}\right) _{j}^{a}=F_{ij}^{a}$ are the only zero
modes of $\Omega _{ab}^{ij}$, and (\emph{ii}) the rank of $\Omega _{ab}^{ij}$
is largest, $\Re \left( \Omega _{ab}^{ij}\right) =2nN-2n$.
\end{thebibliography}
\end{document}